\documentclass[aps,prl,twocolumn,groupedaddress,showpacs]{revtex4}

\newcommand{\eq}{\begin{equation}}
\newcommand{\ee}{\end{equation}}
\newcommand{\eqa}{\begin{eqnarray}}
\newcommand{\eea}{\end{eqnarray}}

\usepackage{graphics}% Include figure files
\usepackage{epsfig}
\usepackage{times}
\usepackage{bm}

\begin{document}
\title{Quantum coherence and Kondo effect in multi quantum dot systems}

\author{Xi Dai }
\affiliation{Department of Physics and Astronomy,
Rutgers University,
Piscataway,
NJ 08854-8019
}

\author{Tai-Kai Ng}
\affiliation{Department of Physics, Hong Kong University of Science and Technology, Clear Water Bay,
Kowloon, Hong Kong}
\date{\today}

\begin{abstract}
The quantum interference effect among coupled identical quantum
dots is studied in the present paper in the limit of strong
intra-dot Coulomb interaction. When the average electron number in
each dot is a fraction of an integer, quantum interference effect
is greatly enhanced because of the sharing of extra electrons by
multiple dots. We show that if the extra electron (hole) number is
one, the low energy effective Hamiltonian can be map into the
2-channel SU(M) Coqblin-Schriffer model, where M is the total dot
number. In particular, for two-dot system with odd number of total
electrons, the model is equivalent to a two channel Kondo problem
with anisotropic coupling between local spin and conduction bands.
The more general situation with (arbitrary) fractional average
electron number in each dot is also discussed. To study the Kondo
effect, we apply the self consistent ladder approximation(SCLA) to
study the electron spectral function for the two-dot system.
Similar method is also used to study the triple-dot system where
we show how quantum coherence manifest itself in the Aharonov-Bohm
(AB) effect.

\end{abstract}
%]
\maketitle
\newpage
Quantum interference effect among coupled quantum dots is becoming
a very important problem because of its possible application in
quantum computing\cite{Loss} and has been studied widely both
theoretically\cite{Golden,Ruzin,Matveev,Izumida} and
experimentally\cite{Waugh,Fujisawa,Oosterkamp}. Similar to single
quantum dot, which are regarded as ``artificial atoms'', the
coupled quantum dots are viewed as ``artificial
molecules''\cite{2dot_review}. The simplest way to model two
coupled quantum dots is to treat the problem as a two-level
system, where only one quantum level in each quantum dot is
considered. When the quantum tunnelling term between the two dots
is increased gradually, a transition from ``ionic bonding'' (where
the electron is localized) to ``covalent bonding'' (where the
electron is sheared by two dots) is expected\cite{2dot_review}. In
the latter case, the two energy levels will split to form bonding
and anti-bonding states. This effect can be detected by photo emission
measurement, where two delta peaks would appear right at the energies of the bonding and anti-bonding states. Unfortunately such a simple model is not applicable to
real quantum dot systems, which contain many energy levels as
well as strong coulomb interaction. In the present paper, we
consider the opposite limit where the energy-level spacing within
each dot is small and treat the energy levels in each dot in the
continuum limit. This approximation is valid for temperature
larger than the level spacing and is much closer to most real
situations.

When M identical quantum dots are coupled by tunnelling processes,
the total electron numbers in the system is a key factor
determining the low energy physics of the system in the presence
of strong coulomb interaction. If the total electron number equals
$M\cdot N$, which is called the commensurate case, the
configuration with the lowest charging energy is precisely $N$
electrons per dot. Any tunnelling processes will cost two times of
the charging energy and thus be suppressed in low temperature.
Therefore the tunnelling term can be treated perturbatively and
the coupling among quantum dots can be viewed as ``ionic
bonding''. However the situation changes completely in the
incommensurate case where the average electron number per dot is
no longer an integer. In the present paper, we will focus on the
simplest situation, where the total electron number equals $M\cdot
N+1$ (extra electron) or $M\cdot N-1$ (extra hole). In this case
we have $M$ degenerate charging states with lowest energy. The
$i^{th}$ state ($i=1,...M$) has $N+1$ ($N-1$ for extra hole)
electrons in the $i^{th}$ dot and $N$ electrons per dot in the
rest of them. Turning on the quantum tunnelling terms will mix
these $M$ degenerate charging states non-perturbatively. The extra
electron can be viewed as a ``valence electron'' of the system and
a ``covalent bond'' is formed by equally sharing the ``valence
electron'' by all the quantum dots in the ground state. This
non-perturbative picture implies that perturbative treatment of
the tunnelling term will break down in low temperature. We shall
show that by projecting the Hamiltonian into the Hilbert space
with the $M$ lowest charging states, which is valid for $k_BT<<$
the charging energy $U$, the $M$-dot problem can be mapped to a
two-channel SU(M) Coqblin-Schriffer model where the physics of the
system can be understood in analogy with Kondo physics. The Kondo
temperature $T_K$ is the energy scale below which quantum
interference between quantum dots becomes important. At
temperature well above $T_k$, quantum coherences between dots are
lost and the charge dynamics between quantum dots are
classic-resistor-like. At temperatures below $T_K$, the quantum
interference effect builds up and a sharp coherence peak appears
in the photo emission spectra, which is similar to the resonance
peak in Kondo problem. By examining the Aharonov-Bohm (AB) effect
in a three-dot system, we show that the coherent peak is
associated with quantum coherent transport and can be interpret as
a quasi-particle peak.

We start with the Hamiltonian for the $M$-dot system with $M\cdot N+1$ electrons,
\begin{equation}
H_{M-dot}=H_{cb}+H_0+H_T
\end{equation}
where
\begin{equation}
H_{cb}=U\sum_{\alpha=1}^M(\hat n_{\alpha}-n_0)^2+V_{gate}\sum_{\alpha=1}^M\hat n_{\alpha}
\end{equation}
is the coulomb blockade term with $n_{\alpha}$ being the particle
number in the $\alpha^{th}$ dot, $n_0$ is the electron number of a
neutral dot, and $V_{gate}$ is the gate voltage applied to the
multi-dot system to tune the chemical potential.
\begin{equation}
H_0=\sum_{\alpha,k,\sigma}\epsilon_{\alpha k \sigma}C_{\alpha k \sigma}^+
C_{\alpha k \sigma}
\end{equation}
and
\begin{equation}
H_T=\sum_{k,k',\sigma,<\alpha\beta>}t_{k,k'}C_{\alpha k \sigma}^+C_{\beta k' \sigma}+H.C.
\end{equation}
are the kinetic energy of each quantum dot and tunnelling term
between them, respectively. We note that similar Hamiltonian has
been used to study the transport properties of coupled double-dot
system by many authors\cite{Golden,Matveev}. Here we include only
the self-capacitance (Coulomb Blockade) term in our interaction
term $H_{cb}$.

We shall map approximately our Hamiltonian to a quantum rotor
Hamiltonian. The mapping is a generalization of the Matveev's
mapping\cite{Matveev2} of the single quantum dot Coulomb Blockade
problem to the Kondo problem, and is valid when the total energy
level number ($N_{total}$) and total number of electrons $N$
satisfy $N_{total},N>>1$.  In
the rotor notation, the Hamiltonian can be written as,
$$
H_{M_dot}=U\sum_{\alpha=1}^M(\hat L_{\alpha}^z)^2+V_{gate}\sum_{\alpha=1}^M\hat L_{\alpha}^z
+\sum_{\alpha,k,\sigma}\epsilon_{\alpha k \sigma}f_{\alpha k \sigma}^+
f_{\alpha k \sigma}
$$
\begin{equation}
+\sum_{k,k',\sigma,<\alpha\beta>}t_{k,k'}f_{\alpha k \sigma}^+f_{\beta k' \sigma}{\hat L^+_\alpha}
{\hat L^-_\beta}+H.C.
\end{equation}

where $\hat L_{\alpha}^z$ is the angular momentum in z axis of a
quantum rotor and represents physically the excess charge on dot
$\alpha$, $\hat L^+$,$\hat L^-$ are the raising and lowering
operator of a quantum rotor. The physical electron operator at dot
$\alpha$ is $ C_{\alpha k \sigma}^+= f_{\alpha k \sigma}^+{\hat
L^+_\alpha}$.  The rotor representation is exact if $M$ local
constrains, $\sum_{l}f^+_{\alpha,l}f_{\alpha,l}=n_0+\hat
L^z_{\alpha}$, for $i=1,...,M$, are introduced to reproduce the
correct Hillbert space. With $N_{total},N>>1$, the above
constrains can be treated in a large-$N$ type expansion. In the
large-$N$ limit, the constraints are satisfied on average by
properly choosing the chemical potential, the total fermion number
on each dot has very small fluctuations.

If $V_{gate}=0$, the ground state without tunnelling term is
unique, corresponding to neutral dots with the fermion levels all
filled up to the Fermi energy. In this case the quantum tunnelling
term can be treated perturbatively because the ground state is
gaped. The quantum coherence between dots is very weak. By
applying proper gate voltage we can shift the chemical potential
such that states with total charge one are the lowest charging
states. In this case we have M degenerate low energy states,
corresponding to the excess electron located at $M$ different
dots. Since we are only interested in the low temperature region
$k_BT<<U$, we can safely neglect all the other charging states and
project the above Hamiltonian to the subspace containing only the
above M charging states. The projected Hamiltonian can be written
as,

\begin{equation}
H_{eff}=\sum_{k,\alpha,\sigma}\epsilon_k f_{k\alpha\sigma}^+f_{k\alpha\sigma}
+{t}\sum_{kk',\alpha\beta,\sigma}f_{k\alpha\sigma}^+f_{k'\beta\sigma}|\alpha><\beta| + H.C.
\label{extra1}
\end{equation}

where $|\alpha>$ represents a state with the extra electron on dot
$\alpha$. The above Hamiltonian can be viewed as a two channel
SU(M) Coqblin-Schriffer model. In double-dot case, it is also
equivalent to two channel anisotropic Kondo
mode\cite{Matveev2,Lebanon}, where the two lowest energy charging
states are equivalent to two pseudo spin states, up and down.
Because of the presence of degenerate states, a perturbation
series in $t_{k,k'}$ will diverge logarithmically at low
temperature, reflecting the set up of quantum interference below
the ``Kondo Temperature'' $T_k$, where the extra electron is
allowed to hop around the quantum dots coherently forming a
``covalent bond'' solid including all $M$ dots.

In the case with more than one extra electrons (holes), we obtain
the following similar low energy effective model following the
same procedure,
$$
H_{eff}=\sum_{k,\alpha,\sigma}\epsilon_k f_{k\alpha\sigma}^+f_{k\alpha\sigma}
$$
\begin{equation}
+{t}\sum_{kk',\alpha\beta,\sigma}f_{k\alpha\sigma}^+f_{k'\beta\sigma}D_{\alpha\beta}
^{\gamma\delta}|\gamma><\delta| + H.C.
\label{general}
\end{equation},
where $D_{\alpha\beta}^{\gamma\delta}=<\gamma|\hat
L^+_{\alpha}\hat L_{\beta}^-|\delta>$ and $<\gamma|$,$<\delta|$
denote all the degenerate charging states with the lowest energy.
Notice that the number of degenerate states increases rapidly with
more dots and extra electrons (holes). It is easy to see that
Eq.\ref{general} reduces to Eq. \ref{extra1} when the extra
electron (hole) is limited to one. We will discuss the properties
of this general model elsewhere.

\begin{figure}
\leavevmode
\epsfxsize7.0cm
\epsfbox{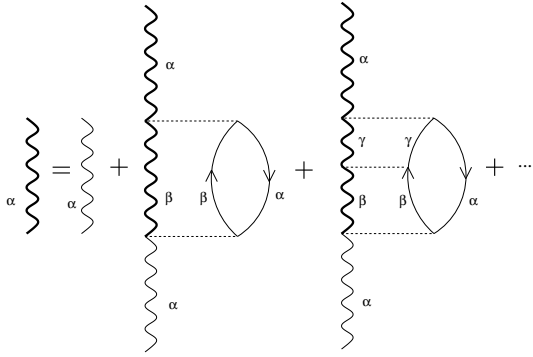}
\caption{\label{fig:fmdiagram}
(Color) The diagrams included in SCLA to calculate the renormalized Green's function
of $\alpha$th charging state. The thin and thick wave lines represent the bare and
full Green's function of the $\alpha$th charging state respectively. The solid lines
represent the occupation number of fermions. And the dashed lines represent the
interaction between fermions and local charging states.
}
\end{figure}

 In the following we shall apply the self consistent ladder approximation (SCLA)
\cite{Lebanon,Maekawa,Bickers} to solve the above effective
Hamiltonian for double- and triple- dots. The SCLA can be viewed
as the generalization of the non-crossing approximation for the
Kondo Hamiltonian and works very well in all temperature range in
the two channel case. Recently the same method has been applied to
the problem of charge fluctuation in a quantum box and the results
are quite satisfactory compared with Bethe Ansatz.\cite{Lebanon}
Using the notation and diagrammatic rule in reference
\cite{Bickers}, the diagrams included by SCLA are shown in
Fig.\ref{fig:fmdiagram}, where the self energy of the charging
state $|m>$ is obtained from the irreducible T-matrix of the
charging states and fermions. The T-matrix is obtained by summing
up all the ladder diagrams using the fully dressed Green's
function of the charging states, which will close the self
consistent integral equations. The SCLA integral equations were
solved under different temperature iteratively and all the details
can be found in reference\cite{Lebanon,Maekawa}. We shall first
examine the electron spectral function in the two-dot system which
can be seen directly in photo emission spectra.

\begin{equation}
G_{\sigma\alpha}(i\nu_n)=\sum_{l,l'}\int_0^{\beta} e^{i\nu_n\tau}
<T L_{\alpha}^-(\tau)f_{l\sigma\alpha}(\tau)L_{\alpha}^+f^+_{l'\sigma\alpha}>d\tau
\end{equation}

\begin{figure}
\leavevmode
\epsfxsize3.0cm\epsfbox{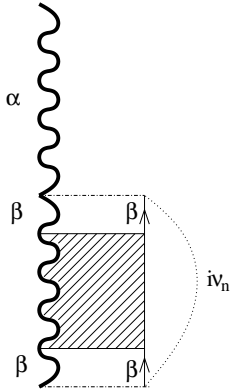}
\caption{\label{fig:gfdiagram}
The diagrams used to calculate physical electron Green's function.The dotted line represents
the frequency carried by the green's function and the dot dashed lines represent the creation
and annihilation of a physical electron.
The shaded area represent all the ladder diagrams in Fig.\ref{fig:fmdiagram}.
}
\end{figure}

\begin{figure}
\leavevmode
\epsfxsize7.0cm\epsfbox{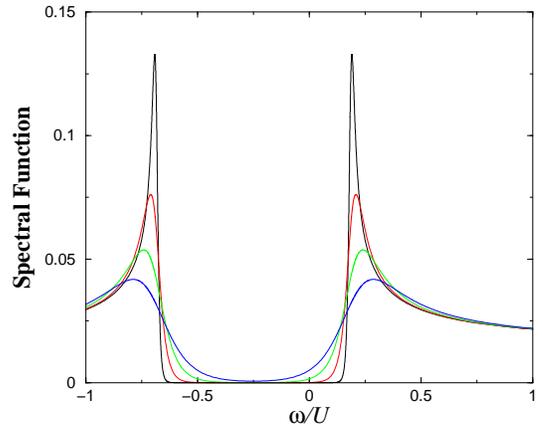}
\caption{\label{fig:A_2dot}
The electronic spectral function for the double-dot system with temperature $T=
0.01,0.05,0.1,0.2$, from the top to bottom.}
\end{figure}

To calculate the above Green's function, we have to include more
charging states other than the ones with the lowest energy. Here
we consider a double-dot system with
$V_{gate}=0.5,U=1.0,N_lt=4.0$, where $N_l$ is the total number of
levels in a single dot. A semi circle density of states with half
band width equals $D=10.0$ is used as the density of states for a
single dot. Following reference \cite{Lebanon} we can estimate the
Kondo temperature by $ T_K=(2D\rho_0t)exp\left
[-{{\pi}\over{4\rho_0t}}\right ] $, which is $0.23$ for the above
parameters. Only the low energy charging states with total charge
number equals $-1,0,1$ are considered to obtain the electron
Green's function.The generalization of SCLA to include more
charging states is quite straightforward. Since the total charge
number is a good quantum number, we can simply apply the SCLA to
the subspace with total charge number equals $1$ and $-1$,
respectively by the above mapping and obtain the renormalized
Green's function for each charging states. In the subspace with
total charge zero, the charging state with the lowest energy is
unique and the Green's function can be obtained by second order
perturbation theory. Using the renormalized Green's function for
the different charging states, the electron Green's function can
be obtained following the diagrammatic rules in
reference\cite{Bickers}. The diagram we used to calculate the
electron Green's function is shown in Fig.\ref{fig:gfdiagram}. The
electron spectral function is shown in Fig.\ref{fig:A_2dot}. In
high temperature, where the quantum interference effects are weak,
a featureless spectra on top of the coulomb gap is found, which
reflects the continuous energy levels inside each quantum dot. As
temperature decreases, sharp resonance peaks which represent the
"covalent bonding" states emerge gradually from the featureless
continuum.

To study the quantum interference effect among the quantum dots
more carefully, we examine the Aharonov-Bohm (AB) effect in the
triple-dot system by connecting two leads to a three-dot system. In
the weak lead-system coupling limit, we can treat the quantum
tunnelling within the triple-dot system by SCLA and
treat the lead-system tunnelling by first order perturbation. The
Master's equation technique\cite{Glazman} is used to calculate the current
flowing through the triple-dot system when a finite voltage is
added between two leads. A magnetic flux enclosed by the dots can
be included by introducing non trivial phase factors to the
tunnelling terms between different quantum dots.

\begin{figure}
\leavevmode
\epsfxsize6.0cm\epsfbox{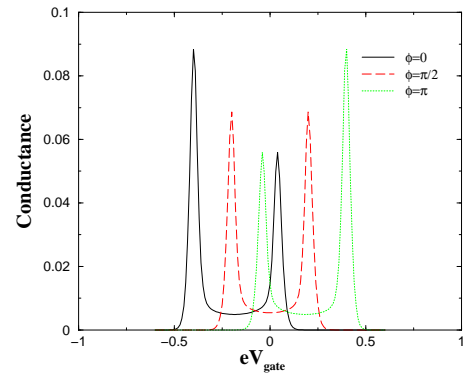}
\caption{\label{fig:cond}
(Color) The zero bias conductance as the function of gate voltage under different magnetic flux.
}
\end{figure}

\begin{figure}
\leavevmode
\epsfxsize6.0cm\epsfbox{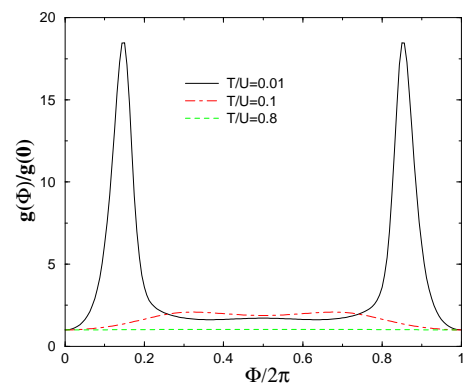}
\caption{\label{fig:mag-cond}
(Color) The normalized DC conductance as the function of magnetic flux $\Phi$.
}
\end{figure}

In Fig.\ref{fig:cond} we plot our results for the zero bias
conductance as the function of gate voltage under different
magnetic flux. The parameters we used here are
$N_lt=2.0$,$U=1.0$,$D=10.0$ and $T=0.01$. We observe that the
presence of the two resonance peaks in the electron spectral
function, corresponding to transitions from charging state $-1$ to
$0$ and from $0$ to $1$, is reflected in conductance measurement.
The resonance peaks move to different gate voltages when magnetic
flux changes, i.e. the quantum coherence manifest itself as a
giant {\em magnetoresistance} effect at low temperature in the
three-dot system which can be observed in transport measurements.
We also plot the renormalized DC conductance as a function of
magnetic flux in Fig.\ref{fig:mag-cond} for gate voltage
$V_g=0.1$, $N_lt=2.0$,$U=1.0$ at different temperatures. At
temperature $T=0.01$ we can clearly see two resonance peaks
corresponding to the resonance between charging states $-1$ to $0$
and $0$ to $+1$. The resonance feature disappears when the
temperature is raised above the Kondo temperature $T_K$ which is
around $0.2$ with our parameters. Notice that $g(\Phi)/g(0)\sim
O(1)$ for temperature $\geq T_K$, consistent with the expectation
that the system behaves like a classical resistor network at $T
\geq T_K$.

 We note that the quantum interference we studied in the present
paper is a many-body effect which is established with the help of strong Coulomb interaction.
The low temperature quantum coherent state is a covalent-bond
solid state described by Kondo-type physics which is possible
because of strong reduction in the number of low-energy charge
excitations. This is contrary to the usual view where Coulomb
interaction is considered as one of the origins of dephasing. It
would be interesting to see how quantum coherence is modified at
low temperature when Coulomb repulsion $U$ decreases.

Summarizing, we show that quantum interference can be set up at
low temperature between small particles where the Coulomb
interaction is strong if the average electron numbers of the
quantum dots are not an integer. The extra electrons behave like the
valence electrons in molecular system to build the "covalent
bonds" among quantum dots. For $M$ coupled identical quantum dots
with one extra charge, the low energy effective Hamiltonian is a
two-channel SU(M) Coqblin-Schrieffe model. In a two-dot system,
the effective Hamiltonian is equivalent to a two channel Kondo
model and quantum coherence will set up at temperatures
lower than an effective "Kondo temperature". We compute the
electronic spectral function using SCLA and sharp quasi particle
peaks are found below the Kondo temperature. For triple-dot system
the low energy effective Hamiltonian is 2-channel SU(3)
Coqblin-Schrieffer model and we can apply the same SCLA method to
the problem. We study transports in the triple-dot
system by the Master's equation technique. A strong AB effect is
found below the "Kondo temperature" $T_K$ indicating the presence of strong
quantum coherence effect.

T.K. Ng thanks Hong Kong Research
Grant Council for support through Grant no. HKUST6142/00P.


\begin{thebibliography}{99}

\bibitem{Loss}
D. Loss and D.P. DiVincenzo, Phys. Rev. A {\bf 57}, 120(1998).

\bibitem{2dot_review}
W.G. van der Wiel {\it et al.}, Rev. Mod. Phys. {\bf 75}, 1 (2003).

\bibitem{zhang_xx}

X.X. Zhang {\it et al.}, Phys. Rev. Lett. {\bf 86}, 5562 (2001).

\bibitem{Golden}
J.M. Golden and B.I. Halperin, Phys. Rev. B {\bf 53}, 3893(1996).

\bibitem{Ruzin}
I.M. Ruzin {\it et al.},Phys. Rev. B {\bf 45}, 13469 (1992).

\bibitem{Matveev}
K.A. Matveev {\it et al.}, Phys. Rev. B {\bf 54}, 5637 (1996).

\bibitem{Izumida}
W. Izumida and O. Sakai, Phys. Rev. B {\bf 62}, 10260 (2000).

\bibitem{Waugh}
F.R. Waugh {\it et al.},Phys. Rev. B {\bf 53}, 1413 (1996);
F.R. Waugh {\it et al.},Phys. Rev. Lett. {\bf 75}, 705 (1995).

\bibitem{Fujisawa}
T. Fujisawa {\it et al.}, Science {\bf 282}, 932 (1998).

\bibitem{Oosterkamp}
T.H. Oosterkamp {\it et al.}, Nature {\bf 395}, 873 (1998).

\bibitem{Matveev2}
K. A. Matveev, Zh.Eksp. Teor. Fiz. {\bf 99}, 1598 (1991)
[Sov. Phys. JETP {\bf 72}, 892 (1991)].

\bibitem{Lebanon}
E. Lebanon {\it et al.}, Phys. Rev. B {\bf 64}, 245338 (2001)

\bibitem{Maekawa}
S. Maekawa {\it et al.}, J. Phys. Soc. Jap. {\bf 54}, 1955 (1985).

\bibitem{Bickers}
N.E. Bickers, Rev. Mod. Phys. {\bf 59}, 845 (1987).

\bibitem{Glazman}
L.I. Glazman {\it et al.}, J. Phys.: Condens. Matter {\bf 1}, 5811 (1989);
I.O. Kulik {\it et al.}, Sov. Phys.-JEPT {\bf 41}, 308 (1975).


\end{thebibliography}
\end{document}